\documentstyle[a4,11pt,menu99,epsfig]{article}
\begin{document}
\setlength{\baselineskip}{2.6ex}

\title{Detection of Pionium with DIRAC}

\author{A. Lanaro\\
{\em CERN, Geneva, Switzerland, and INFN-Laboratori Nazionali di Frascati,
Frascati, Italy} \\
\vspace*{0.3cm}
on behalf of\\
``The DIRAC Collaboration"\\
\vspace*{0.3cm}
B.~Adeva$^o$, L.~Afanasev$^l$, M.~Benayoun$^d$, 
V.~Brekhovskikh$^n$,
G.~Caragheorgheopol$^m$,T.~Cechak$^b$, M.~Chiba$^j$, S.~Constantinescu$^m$,
A.~Doudarev$^l$, D.~Dreossi$^f$, D.~Drijard$^a$, 
M.~Ferro-Luzzi$^a$, T.~Gallas Torreira$^{a,o}$, J.~Gerndt$^b$, R.~Giacomich$^f$,
P.~Gianotti$^e$, F.~Gomez$^o$, A.~Gorin$^n$, O.~Gortchakov$^l$, 
C.~Guaraldo$^e$, M.~Hansroul$^a$, R.~Hosek$^b$,
M.~Iliescu$^{e,m}$, N.~Kalinina$^l$, V.~Karpoukhine$^l$, 
J.~Kluson$^b$, M.~Kobayashi$^g$, P.~Kokkas$^p$, 
V.~Komarov$^l$, A.~Koulikov$^l$, 
A.~Kouptsov$^l$, V.~Krouglov$^l$, L.~Krouglova$^l$, K.-I.~Kuroda$^k$,
A.~Lanaro$^{a,e}$, V.~Lapshine$^n$, R.~Lednicky$^c$, P.~Leruste$^d$, 
P.~Levisandri$^e$, A.~Lopez Aguera$^o$, V.~Lucherini$^e$, 
T.~Maki$^i$, I.~Manuilov$^n$, L.~Montanet$^a$, 
J.-L.~Narjoux$^d$, L.~Nemenov$^{a,l}$, M.~Nikitin$^l$, 
T.~Nunez Pardo$^o$, K.~Okada$^h$, V.~Olchevskii$^l$, 
A.~Pazos$^o$, M.~Pentia$^m$, A.~Penzo$^f$, J.-M.~Perreau$^a$, 
C.~Petrascu$^{e,m}$, M.~Plo$^o$, T.~Ponta$^m$, D.~Pop$^m$, 
A.~Riazantsev$^n$, J.M.~Rodriguez$^o$,
A.~Rodriguez Fernandez$^o$, V.~Rykaline$^n$,  
C.~Santamarina$^o$, J.~Schacher$^q$, A.~Sidorov$^n$, J.~Smolik$^c$,
F.~Takeutchi$^h$, A.~Tarasov$^l$, L.~Tauscher$^p$, S.~Trousov$^l$, 
P.~Vazquez$^o$, S.~Vlachos$^p$, V.~Yazkov$^l$, Y.~Yoshimura$^g$, P.~Zrelov$^l$
\vspace*{0.3cm}\\
$^a$ CERN, Geneva, Switzerland \\
$^b$ Czech Technical University, Prague, Czech Republic\\
$^c$ Prague University, Czech Republic\\
$^d$ LPNHE des Universites Paris VI/VII, IN2P3-CNRS, France\\
$^e$ INFN - Laboratori Nazionali di Frascati, Frascati, Italy\\
$^f$ Trieste University and  INFN-Trieste, Italy\\
$^g$ KEK, Tsukuba, Japan \\
$^h$ Kyoto Sangyou University, Japan\\
$^i$ UOEH-Kyushu, Japan\\
$^j$  Tokyo Metropolitan University, Japan\\
$^k$ Waseda University, Japan\\
$^l$ JINR Dubna, Russia\\
$^m$ National Institute for 
Physics and Nuclear Engineering IFIN-HH, Bucharest, Romania \\
$^n$ IHEP Protvino, Russia\\
$^o$ Santiago de Compostela University, Spain \\
$^p$ Basel University, Switzerland \\
$^q$ Bern University, Switzerland 
}

\maketitle

\begin{abstract}
\setlength{\baselineskip}{2.6ex}

The aim of the DIRAC experiment at CERN is to provide an accurate
determination of S-wave $\pi\pi$ scattering lengths
from the measurement of the lifetime of the $\pi^+\pi^-$ 
atom. The measurement will be done with precision comparable to the level
of accuracy of theoretical predictions, formulated in
the context of Chiral Perturbation Theory. Therefore, 
the understanding of chiral symmetry breaking of QCD
will be submitted to a stringent test.

\end{abstract}

\setlength{\baselineskip}{2.6ex}

\section*{INTRODUCTION}
The low-energy dynamics of strongly interacting hadrons is under the
domain of non-perturbative QCD, or QCD in the confinement region. At present, 
low energy 
pion-pion scattering is still an unresolved problem in the context of
QCD. However, the approach based on effective chiral Lagrangian has been
able to provide accurate predictions on the dynamics of light hadron
interactions [1]. In particular, Chiral Perturbation Theory (CHPT) allows
to predict the S-wave $\pi\pi$ scattering lengths at the level of few percent
[2]. Available experimental results, on their side, are much less 
accurate than theoretical predictions, both because of large experimental 
uncertainty and, in some cases, unresolved model dependency [3].

The DIRAC experiment aims at a model independent measurement of the difference
$\Delta$ between the isoscalar $a_0$ and isotensor $a_2$ S-wave $\pi\pi$
scattering lengths with $5\%$ precision, by measuring the lifetime of the 
pionium ground state with $10\%$ precision.
 
\section*{PIONIUM}

Pionium ($A_{2\pi}$) is a Coulomb weakly-bound system of a $\pi^+$ and 
a $\pi^-$, whose lifetime
is dominated by the charge-exchange process to two neutral pions. The Bohr
radius is 387 fm, the Bohr momentum 0.5 MeV/c, and the binding 
energy 1.86 keV. The decay probability is proportional to the atom wave
function squared at zero pion separation and to the square of $\Delta=a_0-a_2$.
Using the values of scattering lengths predicted by CHPT, the lifetime of the 
$\pi^+\pi^-$ atom in the ground state is predicted to be 3.25$\times$10$^{-15}$ s 
[2]. 

\subsection*{Production of $A_{2\pi}$}

In DIRAC, $\pi^+\pi^-$ atoms are formed by the interaction of 24 GeV/c protons 
with nuclei in thin targets [4]. If two final state pions have a small relative 
momentum in their system ($q\sim~1MeV/c$), and are much closer than the Bohr
radius, then 
the $A_{2\pi}$ production probability, due to the high overlap, is large.
Such pions originate from short-lived sources (like $\rho$ and
$\omega$), but not from long-lived ($\eta$, $K^0_s$), because in the latter case
the two-pion separation is larger than the Bohr radius. The production 
probability for $A_{2\pi}$ can then be calculated using the double inclusive
production cross section for $\pi^+\pi^-$ pairs from short-lived sources,
excluding Coulomb interaction in the final state [5]. Evidence for $A_{2\pi}$
production was reported in a previous experiment [6].

\subsection*{Fate of $A_{2\pi}$}

Pionium travelling in matter can dissociate or break up into
a pair of oppositely charged pions with small relative momentum ($q<3$ MeV/c) 
and hence with very small angular divergence ($\theta<0.3$ mrad). This process 
competes with the charge-exchange reaction or decay, if the target material is 
dense so that the atomic 
interaction length is similar to the typical decay length of a few GeV/c 
dimeson atom (a few tens of microns). In a 100$\mu$m Ni foil, for example, the
$A_{2\pi}$ breakup probability ($\sim~47\%$) becomes larger than the 
annihilation probability ($\sim~38\%$).
This breakup probability depends on the target nucleus charge Z, the target
thickness, the $A_{2\pi}$ momentum, and on the $A_{2\pi}$ lifetime [5,7].

\subsection*{Measurement of the $A_{2\pi}$ lifetime}

For a target material of a given thickness, the breakup probability for pionium
can be experimentally determined from the measured ratio of the number of
dissociated atoms ($n_A$) to the calculated number of produced $A_{2\pi}$ 
($N_A$). Thus, by comparison with the theoretical value, known at the 1$\%$
level, the $A_{2\pi}$ lifetime can be determined.

The number $n_A$ of detected ``atomic pairs" is obtained from the experimental
distribution of relative momenta $q$ for pairs of oppositely charged pions.
It is however necessary to subtract a background contribution, arising mainly 
from Coulomb-correlated pions pairs in the $q$ region, where the $A_{2\pi}$
signal is prominent ($q<2$ MeV/c). The low-$q$ background contribution is
obtained with an extrapolation procedure using the shape of the accidental pair
$q$-distribution recorded in the region $q>3$ MeV/c, taking into account e.m. 
and strong $\pi^+\pi^-$ final state interactions [7].

From the measured ratio $n_A/N_A$ a value for the $A_{2\pi}$ ground state
lifetime can be extracted and, hence, a value for $\Delta=|a_0-a_2|$.
    
\section*{THE EXPERIMENTAL APPARATUS}

The DIRAC experimental apparatus (Fig.~1) [4], devoted to the detection 
of charged pion pairs, was installed and commissioned in 1998 at the
ZT8 beam area of the PS East Hall at CERN. After a calibration run at the
end of 1998, DIRAC has been collecting data since the summer of 1999.

The primary PS proton beam of 24 GeV/c nominal momentum struck the DIRAC
target. The non-interacting beam travels below the secondary
particle channel (tilted upwards at 5.7$^o$ with respect to the proton beam), 
until it is absorbed by a catcher downstream of the setup.
Downstream the experimental target, secondary particles travel across the 
following
detectors: three planes of Micro-Strip Gas Chambers (MSGC) and two orthogonal 
stacks of scintillating fibers (Scintillating Fiber Detector SFD) to provide 
tracking information upstream of the spectrometer magnet; two planes of vertical scintillator slabs 
(Ionization Hodoscope IH) to detect the particle energy loss. Downstream the 
IH, the secondary beam enters a vacuum channel extending through the poles of
the spectrometer magnet of 2.3 Tm bending power in the tilted
horizontal plane. Downstream the analyzing magnet, the setup splits into two
arms (inclined by 5.7$^o$ in the vertical plane, and open by $\pm19^o$ in the
horizontal plane) equipped with a set of identical detectors: 14 drift 
chamber (DC) planes, one plane of vertical scintillating strips (Vertical
Hodoscope VH) and one of horizontal strips (Horizontal Hodoscope HH) for
tracking purposes downstream of the magnet; furthermore, a N$_2$ gas-threshold 
Cherenkov counter (CH), 
a Pre-Shower Detector (PS), consisting of Pb converter plates and of 
vertical scintillator slabs, and a Muon counter (MU), consisting of an array of  
vertical scintillator elements placed behind a block of iron absorber, with the
aim of performing particle identification at the trigger or offline levels.

\parbox{8cm}{
\begin{center}
\epsfig{figure=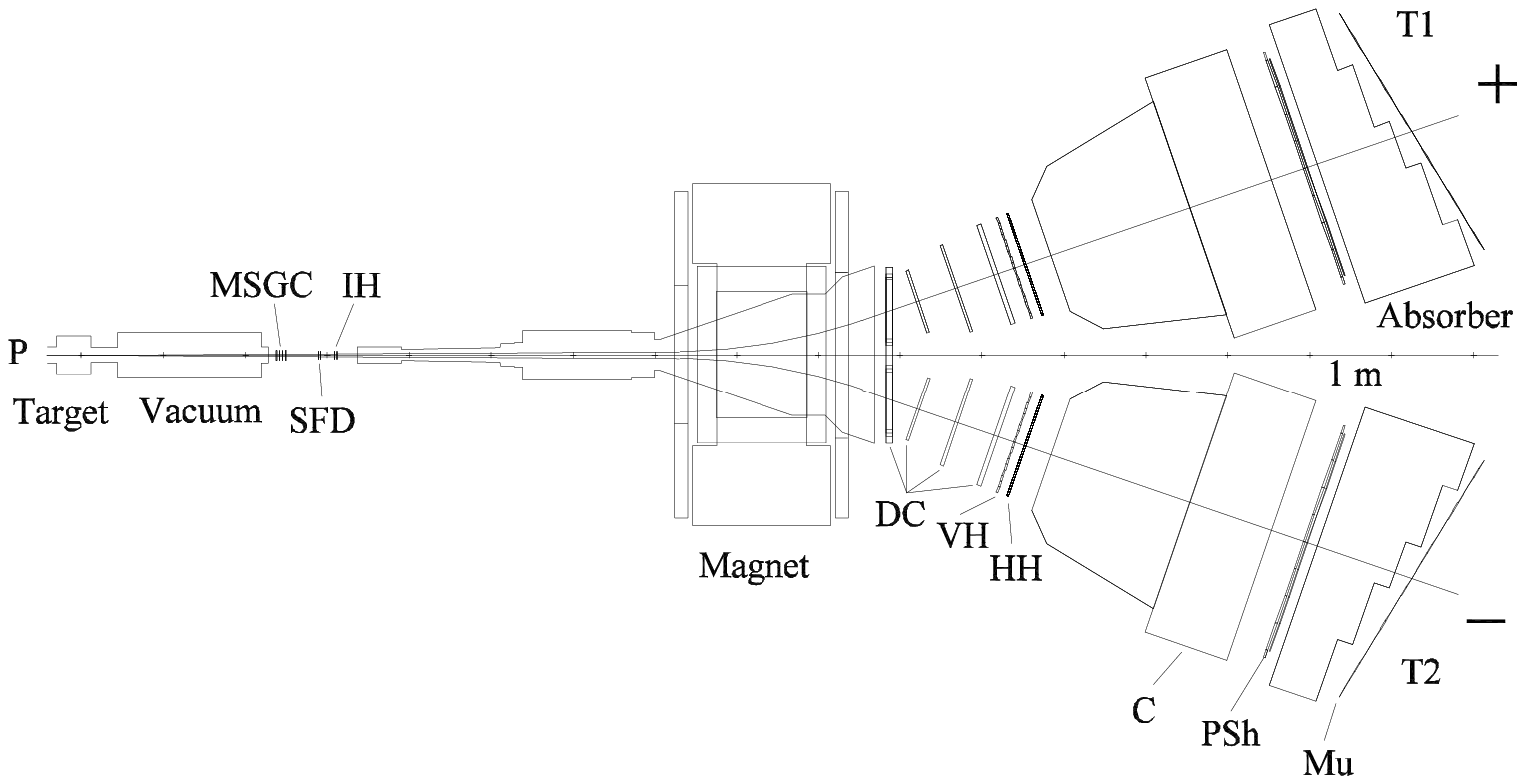,width=8cm,height=6.cm}
\end{center}}
\parbox{7cm}{\vspace*{4.0cm}
\noindent
\parbox{5.2cm}
{\small \setlength{\baselineskip}{2.6ex} Fig.~1. The DIRAC experimental
apparatus.}} 
      
A multi-level trigger was designed to reduce the secondary particles rate 
to a level
manageable by the data acquisition system, and to yield the most favorable
signal-to-noise ratio, by selecting pion pairs with low relative momentum in the
pair system (or small opening angle and equal energies in the lab system) and 
by recording a sufficiently large number
of accidental pairs for the offline analysis. An incoming flux of 
$\sim$10$^{11}$ protons/s would produce a rate of secondaries of about 
3$\times$10$^6$/s and 1.5$\times$10$^6$/s 
in the upstream and downstream detectors, respectively. At the trigger level 
this rate is reduced to about 2$\times$10$^3$/s, with an average event size of about
0.75 Kbytes. With the 95$\mu$m thin Ni target, the expected average $A_{2\pi}$ 
yield in the 
geometrical and momentum setup acceptance is $\sim~0.7\times10^{-3}$/s,
equivalent to a total number of $\sim~10^{13}$ protons on target to produce 
one dimeson atom. 

\section*{RESULTS FROM FIRST DATA TAKING}

A preliminary analysis was performed on a sample of data (Ni target) collected 
during this summer. The sample consisted of about 10$^7$ events,
corresponding to $\sim~$1/3 of the statistics, 
accumulated in a 3-week run period. The data analysis was mostly dedicated 
to the calibration of individual detectors and to the tuning of  
reconstruction algorithms. However, some general features of the
apparatus response were investigated, and some results will be presented 
hereafter.

Figure~2 shows the time difference between hit slabs in the 
left and right
vertical hodoscopes for events with one single track reconstructed in each
detector arm. Within a trigger window of 45 ns, one observes the peak of 
``on-time" hits associated to correlated particles, over the background from 
accidental hits. The width of the correlated-pair events yields the time 
resolution
of the hodoscope ($\sigma\sim~420$ ps at the time of measurement, recently 
improved to $\sim~250$ ps). The asymmetry on the right of the coincidence peak
is due to admixture of protons in the ``$\pi^+$" sample, thus corresponding to 
events of the type $\pi^-p$.

\parbox{8cm}{
\begin{center}
\epsfig{figure=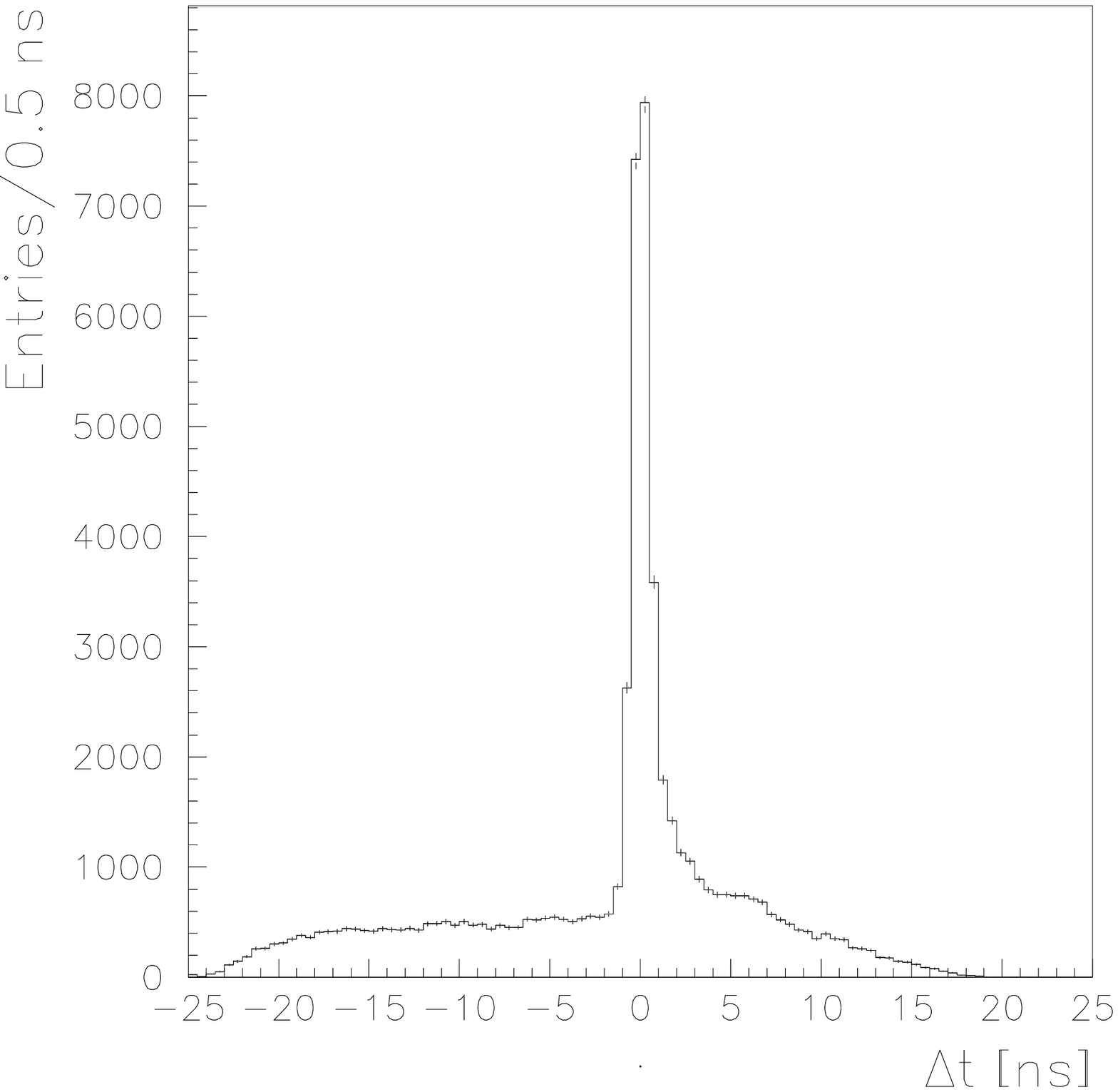,width=6cm,height=6cm}
\end{center}}
\parbox{7cm}{\vspace*{4.0cm}
\noindent
\parbox{5.2cm}
{\small \setlength{\baselineskip}{2.6ex} Fig.~2. Time 
difference between left and right VH scintillator slabs, hit by particles.}}

Such a contamination sample can be isolated by the time-of-flight 
measurement along the path from  the target to the hodoscope. The discrimination
between $\pi^-\pi^+$ and $\pi^-p$ pairs is effective for momenta of positively
charged particles below 4.5 GeV/c. This is shown in Fig.~3, where the 
laboratory momentum
of the positive particle in the pair is shown as a function of the arrival-time 
difference in the vertical hodoscope. The spectrometer single particle 
momentum acceptance is within the range 1.3 to 7.0 GeV/c.

In Fig.~4, the distribution of the longitudinal component ($q_L$) 
of the relative momentum in the pair system is shown 
for two samples of events: those (Fig.~4a) occurring with time 
differences close to zero (real coincidence plus admixture of accidental pairs), 
associated to free pairs with and without final state interaction; and those 
(Fig.~4b) occurring at time differences far from the peak of 
correlated pairs (only accidental pairs).

\parbox{8cm}{
\begin{center}
\epsfig{figure=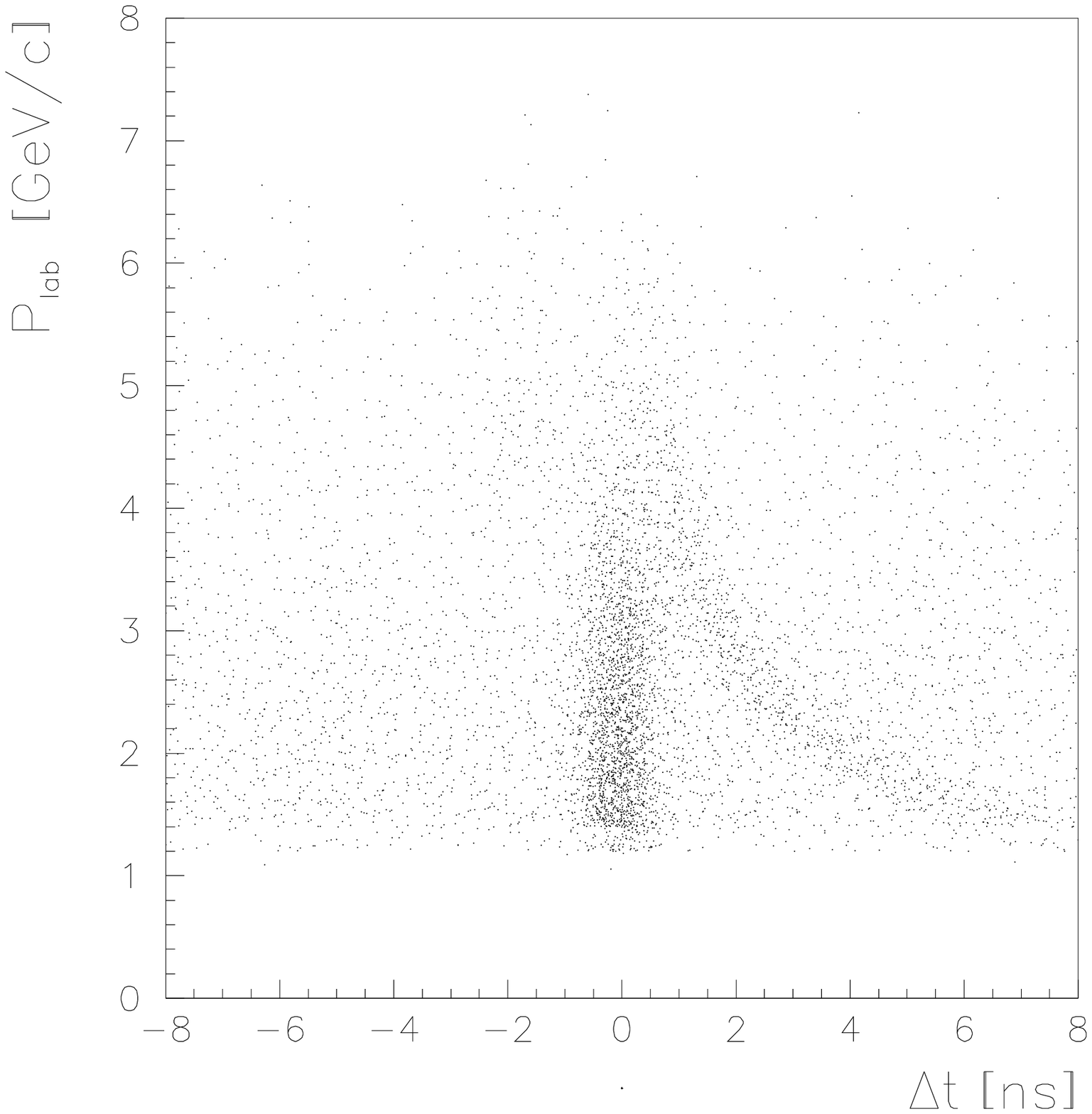,width=6cm,height=6cm}
\end{center}}
\parbox{7cm}{\vspace*{4.0cm}
\noindent
\parbox{5.2cm}
{\small \setlength{\baselineskip}{2.6ex} Fig.~3. Momentum of
positive particle as a function of time difference between left and right 
hit slabs of the vertical hodoscopes.}}

Finally (Fig.~4c), the $q_L$ distribution of correlated pion pairs is obtained 
from the difference between the distributions of Fig.~4a and~4b, 
taking into account the relative normalization factor. The distributions in 
Fig.~4 were obtained from a sample of two-track events, preselected
with momentum of the positive particle less than 4.5 GeV/c, to reject unresolved
$\pi^-p$ pairs, and with transverse component ($q_T$) of the relative momentum below 
4 MeV/c, to increase the fraction of low relative momentum pairs.
For values of $q_L$ corresponding to correlated pairs ($|q_L|<10$ MeV/c) the 
production
cross section of Coulomb pairs is enhanced with respect to the cross section 
of non-Coulomb pairs: Coulomb attraction in the final state is responsible 
for the peak in the $q_L$ distribution (Fig.~4a and~4c) at small $q_L$.
 
A preliminary estimate of the number of pairs associated to $A_{2\pi}$ breakup 
results in a contribution of about 100 ``atomic pairs" in the region $|q_L|<2$ 
MeV/c.

\parbox{8cm}{
\begin{center}
\epsfig{figure=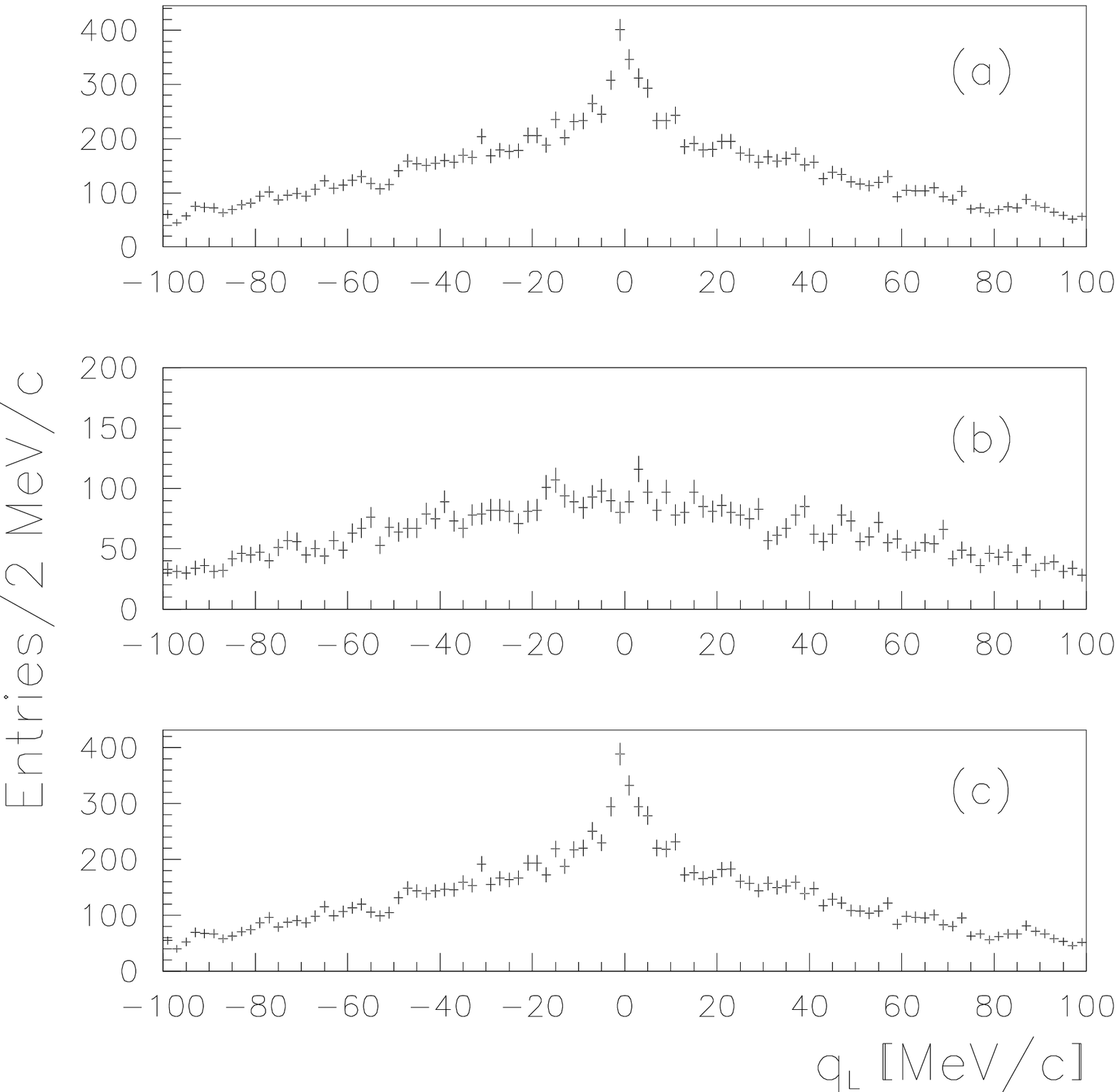,width=6cm,height=6cm}
\end{center}}
\parbox{7cm}{\vspace*{3.8cm}
\noindent
\parbox{5.2cm}
{\small \setlength{\baselineskip}{2.6ex} Fig.~4. Distribution of the
longitudinal component of the relative momentum for: (a) time-correlated pairs;
(b) accidental pairs; (c) spectrum of time-correlated minus accidental pairs.}}

The reconstruction of Coulomb-correlated $\pi^+\pi^-$ pairs is sensitive to the
precision of the setup alignment. Any misalignment of the tracking system
in one arm relative to the other arm would generate asymmetrical errors on the
reconstructed momenta.
This would lead to a systematic shift and
additional spread of the Coulomb enhanced peak in the $q_L$
distribution. The mean value of the Coulomb peak is 0.1 MeV/c, well within the 
accepted tolerances.

When reconstructed momenta of oppositely charged particles are symmetrically
overestimated or underestimated then a calibration using detected resonances is
adequate. This is done by reconstructing the effective mass of $\pi^-p$ pairs, also
detected in the spectrometer. Figure~5 shows the invariant mass distribution
of correlated $\pi^-p$ pairs with proton momentum $>$3 GeV/c.

\parbox{8cm}{
\begin{center}
\epsfig{figure=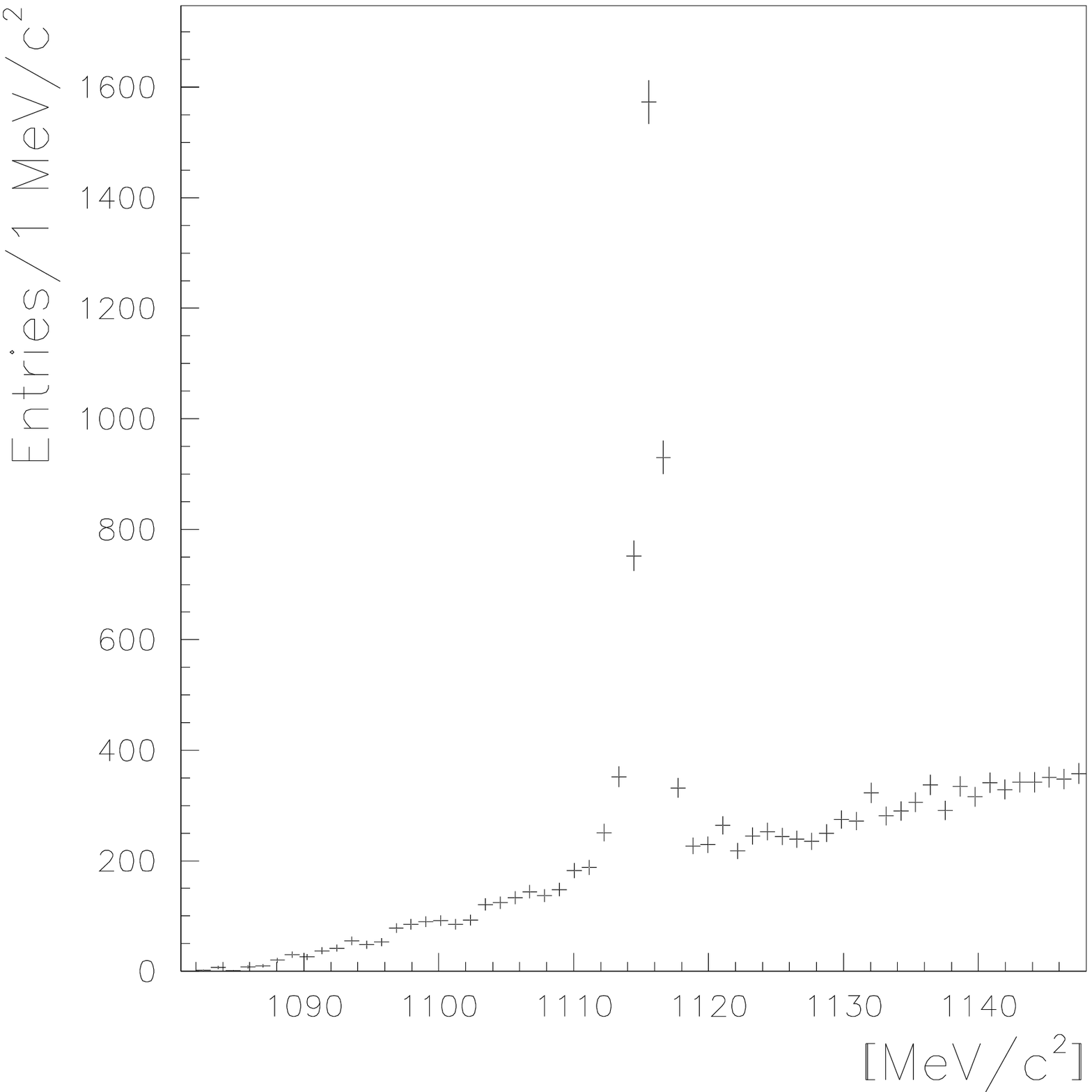,width=6cm,height=6cm}
\end{center}}
\parbox{7cm}{\vspace*{5.0cm}
\noindent
\parbox{5.2cm}
{\small \setlength{\baselineskip}{2.6ex} Fig.~5. Invariant mass of reconstructed
$\pi^-p$ pairs.}}

A clear signal at the nominal $\Lambda$ mass is observed. Such events originate 
from a few GeV/c $\Lambda$, with the decaying 
proton emitted backward and the pion emitted forward in the $\Lambda$ 
system, and both decay particles characterized by small transverse 
momenta. The experimental mean value and standard deviation of the mass peak 
are 1115.60 and
0.92 MeV/c$^2$, respectively. These mass parameter values suggest an excellent
calibration of the momentum scale, with accuracy in momentum reconstruction
better than 0.5$\%$ in the kinematic range of detected $\Lambda$ decays, and
the absence of errors in the telescope alignment, which otherwise
would cause a displacement of the $\Lambda$ mass peak value.

\section*{CONCLUSION}

The DIRAC experiment has begun to collect data this year. A preliminary
investigation of the apparatus performances demonstrates its full capability
to pursue the foreseen experimental program. Improvements to the hardware as 
well as software tools have already been implemented in the second run
period, currently in progress. These will certainly result in better
quality of the data and will contribute to the aimed measurement precision
of the pionium lifetime.

\bibliographystyle{unsrt}

\begin{thebibliography}{99}

\bibitem{1} S.Weinberg, Phys.\ Rev.\ {\bf 166}, 1569 (1968);\\
S.Weinberg, Physica {\bf 96A}, 327 (1979);\\
H.Leutwyler, in  Proceedings of XXVI Conf. on High Energy Physics, {\bf No.272}, 
Dallas, 1992 (AIP, New York, 1993) p.\ 185;\\
J.Gasser and H.Leutwyler, Phys.\ Lett.\ {\bf B125}, 327 (1983);\\
J.Gasser and H.Leutwyler, Nucl.\ Phys.\ {\bf B250}, 465, 517, 539 (1985);\\
J.Stern, H.Sazdjian and N.H.Fuchs, Phys.\ Rev.\ {\bf D43}, 3814 (1993);\\
M.Knecht et al., Nucl.\ Phys.\ {\bf B455}, 513 (1995).

\bibitem{2} J.Bijnens et al.,Phys.\ Lett.\ {\bf B374}, 210 (1996);\\ 
 D.Eiras, and J.Soto, ``Effective Field Theory Approach to Pionium,"
 CERN hep-ph/9905543 v2 (1999) and this Proceedings;\\ 
 V. Lyubovitskij, ``Hadronic Atoms in QCD," this Proceedings;\\
 A.Rusetsky, ``Isospin Breaking Effects in Bound State Observables," this 
 Proceedings.

\bibitem{3} L.Rosselet et al., Phys.\ Rev.\ {\bf D15}, 574 (1977);\\
C.D.Froggat and J.L.Petersen, Nucl.\ Phys.\ {\bf B129}, 89 (1977);\\
W.Ochs, Max Planck Inst. prep. MPI-Ph/Ph 91-35, (1991);\\
M.Kermani et al., Phys.\ Rev.\ {\bf C58}, 3431 (1998).

\bibitem{4} B.Adeva et al., ``Lifetime Measurement of $\pi^+\pi^-$ Atoms to
Test Low Energy QCD Predictions," Proposal to the SPSLC, CERN/SPSLC 95-1,
SPSLC/P 284, (1994).

\bibitem{5} L.L.Nemenov, Yad.\ Fiz.\ {\bf 41}, 980 (1985);\\
O.E.Gorchakov et al., Yad.\ Fiz.\ {\bf 59}, 2015 (1996).

\bibitem{6} L.G.Afanasyev et al., Phys.\ Lett.\ {\bf B308}, 200 (1993);
Phys.\ Lett.\ {\bf B338}, 478 (1994); Yad.\ Fiz.\ {\bf 60}, 1049 (1996).

\bibitem{7} L.G.Afanasyev, JINR E2-91-578, Dubna, (1991);\\  
L.G.Afanasyev and A.V.Tarasov, JINR E4-93-293 Dubna, (1993).

\end{thebibliography}

\end{document}